\documentclass{iopart}

\usepackage{bm}
\usepackage{color}
\usepackage{graphicx}

\begin{document}

\title[Knotted light, helicity, and the flow of null electromagnetic fields]{Linked and knotted beams of light, conservation of helicity and the flow of null electromagnetic fields}
\author{William T.M. Irvine}

\address{James Frack Institute, University of Chicago, Chicago, IL 60637, USA}
\ead{wtmirvine@uchicago.edu}
\begin{abstract}
Maxwell's equations allow for some remarkable solutions consisting of pulsed beams of light which have  linked and knotted field lines. The preservation of the  topological structure of the field lines in these solutions has  previously been ascribed to the fact that the electric and magnetic helicity, a measure of the degree of   linking  and knotting between field lines, are conserved. Here we show that the elegant evolution of the field is due to the stricter condition that the electric and magnetic fields be everywhere orthogonal. The field lines then satisfy a `frozen field' condition and  evolve as if they were unbreakable filaments embedded in a fluid. The preservation of the orthogonality of the electric and magnetic field lines is guaranteed for null, shear-free fields such as the ones considered here by a theorem of Robinson.  We calculate the flow field of a particular solution and find it to have the form of a Hopf fibration moving at the speed of light in a direction opposite to the propagation of the pulsed light beam, a familiar structure in this type of solution. The difference between smooth evolution of individual field lines and conservation of electric and magnetic helicity is illustrated by considering a further example in which the helicities are  conserved, but the field lines are not everywhere orthogonal. The field line configuration at time $t=0$ corresponds to a nested family of torus knots but unravels upon evolution.

\end{abstract}


\maketitle

\section{Introduction}

Light in free space provides an ideal playground for the investigation of geometric and topological structures in fields~\cite{Nye:1974p1870,Allen:1992p1869,Nye:1987p1871}, with linked and knotted structures receiving growing attention of late~\cite{Berry:2001p5,Leach:2004p1864,Dennis:2009p1874,Dennis:2011p1698,Irvine:2008p1699}. Examples of loops, knots and links  have been considered in different `degrees of freedom' of light fields, such as paraxial vortices,  the phase of  Riemann-Siberstein  vectors and field lines; in this article we focus on the rules that govern the evolution of linked and knotted field lines.

To motivate and guide our investigation we consider  a striking solution to Maxwell's equations:  a beam of light whose electric(magnetic) field lines are all closed loops   with any two electric(magnetic) field lines linked to each other. This solution, derived in the context of a topological model for electromagnetism  by Ra\~nada~\cite{Ranada:1989p1769,ref3,ref4,ref5}, studied    by the author in Ref.~\cite{Irvine:2008p1699}  was recently shown by Besieris and Shaarawi~\cite{Besieris:2009p1739} to be equivalent to solutions derived previously by Robinson and Trautman~\cite{ROBINSON:1961p1773,Trautman:1977p1705}, studied and generalized  by Bialynicki-Birula~\cite{BialynickiBirula:2004p1740}. 
A simple expression for the fields due to Ra\~nada is:
\begin{equation}
\bm{B}  = \frac{1}{4\pi i}  \frac{\bm{\nabla}\eta\times\bm{\nabla}\bar{\eta}}{(1+\bar{\eta} \eta)^2} \qquad ; \qquad \bm{E} =  \frac{1}{4\pi i} \frac{\bm{\nabla}\zeta\times\bm{\nabla}\bar{\zeta}}{(1+\bar{\zeta} \zeta)^2},
\end{equation}
\begin{equation}
\zeta(x,y,z,t)=\frac{(Ax+ty)+i(Az+t(A-1))}{(tx-Ay)+i(A(A-1)-tz)},
\label{eq:gop}
\end{equation}
\begin{equation}
\eta(x,y,z,t)=\frac{(Az+t(A-1))+i(tx-Ay)}{(Ax+ty)+i(A(A-1)-tz)},
\label{eq:gap}
\end{equation}
where $A=\frac{1}{2}(x^2+y^2+z^2-t^2+1)$, and $x,y,z,t$ are dimensionless multiples of a length scale  $a$.

\begin{figure}
\begin{center}
\includegraphics[width=0.95\columnwidth]{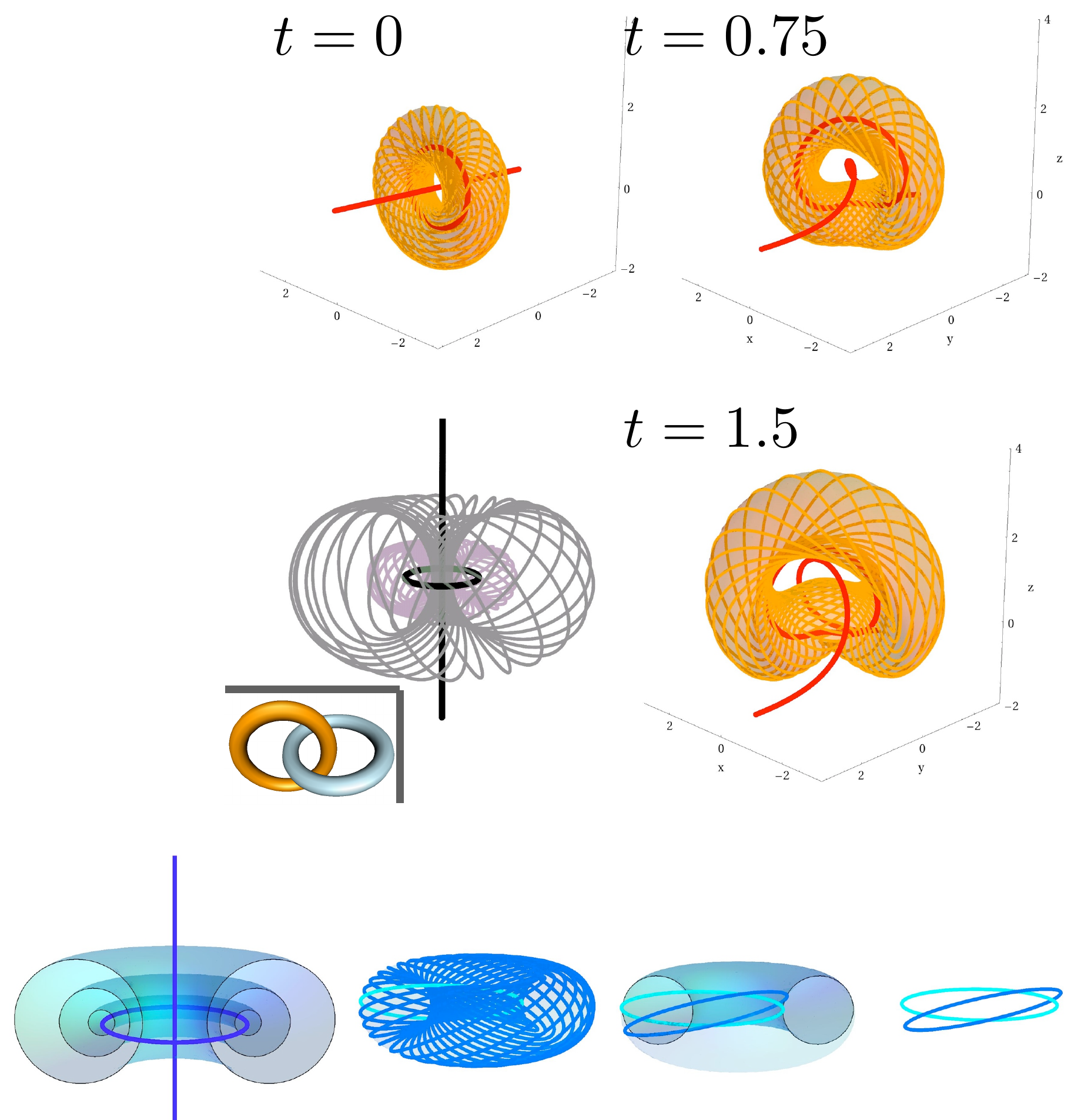}
\caption{An electromagnetic field with linked field lines.  The field lines have 
the structure of the  Hopf fibration, whose construction is shown in blue. 
 The Hopf fibration can be seen as a collection of disjoint circles that fill space with the  property that any two such circles are linked to each other. This remarkable structure can be built by first foliating space with tori of different sizes,  enclosed inside each other like Russian dolls; and subsequently breaking each torus up into a set of circles that wrap once around each circumference of the torus. Since each circle wraps once around each circumference of the torus on whose surface it lives, any two such circles on the same torus will be linked to each other, and, for the same reason, any pair of circles - each from a different torus - will also be linked to each other. To represent such a structure it is convenient to draw the smallest torus (a circle at the center of the nested set),  the infinitely big torus (represented by a straight line through the middle of the nested set) and a set of circles from one or two of the intermediate nested tori. The electric field lines (orange with smallest/largest torus in red) and the Poynting field lines (gray with smallest/largest torus in black) are shown at time $t = 0$, and representative sets of the electric field lines at subsequent times.  The electric field lines have the appearance of filaments that have a continuous identity in time and evolve by stretching and deforming, preserving the topological structure of the field. }
\label{fig:fig1}
\end{center}
\end{figure}

At time $t=0$, the linked field lines are arranged in a structure   known as the  Hopf fibration~\cite{hophinphys,elemhopf}, a collection of disjoint circles that fill space with the  property that any two such circles are linked to each other (See Figure 1, blue). This remarkable structure can be built by first foliating space with tori of different sizes,  enclosed inside each other like russian dolls; and subsequently breaking each torus up into a set of circles that wrap once around each circumference of the torus. Since each circle wraps once around each circumference of the torus on whose surface it lives, any two such circles on the same torus will be linked to each other, and, for the same reason, any pair of circles - each from a different torus - will also be linked to each other. To represent such a structure it is convenient to draw the smallest torus (a circle at the center of the nested set),  the infinitely big torus (represented by a straight line through the middle of the nested set) and a set of circles from one or two of the intermediate nested tori. Figure~\ref{fig:fig1}, shows the electric field lines (orange with smallest/largest torus in red) and the Poynting field lines (gray with smallest/largest torus in black) at time $t=0$, and representative sets of the electric field lines at subsequent times.  Time evolution not only  preserves their topological structure but gives them the appearance of filaments that have a continuous identity in time, and evolve by stretching and deforming.  Natural questions, addressed here, are {\it is this indeed true?  and if so why? }
 
The invariance of topological structure in the fields shown in Fig.~\ref{fig:fig1} has previously been associated~\cite{Ranada:1989p1769} with the conservation of electric and magnetic helicities: $h_{e}$, $h_{m}$ which are a measure of the average linking and knotting of electric/magnetic field lines. Using the  `frozen field' condition~\cite{Newcomb:1958p1003} and applying it to this free-space solution, here we point out that although intimately related to conservation of electric or magnetic helicity, what guarantees such a smooth evolution is in fact a more stringent condition, also satisfied by  the field, which is that the electric and magnetic field are everywhere perpendicular $\bm{E}\cdot \bm{B}=0$.   
 
Indeed for a free-space field satisfying $\bm{E}\cdot \bm{B}=0$, the field lines evolve as if they are stretchable filaments,  transported by a velocity field $\bm{v} = (\bm{E}\times \bm{B})/(\bm{B}\cdot \bm{B})$~\cite{Newcomb:1958p1003}. We calculate this velocity field explicitly for the solution of Fig.~1 and  find it  to have the form of  a Hopf fibration that, without deformation, (unlike the electric and magnetic field structures which deform) moves along the $z-$axis at the speed of light.  This structure corresponds to the core of the  congruence of Robinson~\cite{ROBINSON:1961p1773,robcon}. The field is null ($\bm{E}\cdot \bm{B}=0$ and $\bm{E}\cdot \bm{E}=\bm{B}\cdot \bm{B}$), and  the condition $\bm{E}\cdot \bm{B}=0$ is guaranteed by the shear-free evolution of the field~\cite{ROBINSON:1961p1773,Bampi:1978p1782}. To clarify the difference between smooth evolution of individual field lines and conservation of helicity,  we  consider  a striking case of a field  whose helicity is conserved, but whose field lines do not satisfy everywhere $\bm{E}\cdot \bm{B} =0$. The field lines, that are initially knotted and linked in a striking configuration, break up and change topology  upon evolution.

\section{Electric/magnetic helicity and knottedness}

An electromagnetic field has an infinite number of electric(magnetic) field lines. To quantify how linked or knotted they are,  it is common to use an average measure, the  electric(magnetic) helicity $h_{e(m)}$. $h_{m}$ is given by:
\begin{equation}
h_m = \int  \mathrm{d}^3 x \ \bm{A}(x) \cdot \bm{B}(x), 
\label{eq:eq1}
\end{equation}
where $\bm{B} = \bm{\nabla} \times \bm{A}$. A similar expression, with $\bm{B}$ replaced by $\bm{E}$ and $\bm{A}$ by a field $\bm{C}$ satisfying $\bm{E}= \bm{\nabla} \times \bm{C}$ gives  the electric field helicity $h_{e}$.
$h_{e(m)}$  can be understood as an average measure of how much the field lines are  knotted and linked~\cite{Moffatt:1969p1448}. Following Berger~\cite{Berger:1999p1722},  we first break up the field in to $N$ magnetic flux tubes each with its flux $\phi_i$. For each flux tube pair $i,j$ we then compute the linking number $L_{\ij}$. For a pair of flux tubes,  $L_{ij}$ measures how linked they are to each other and for a single curve  $L_{ii}$ is a measure of knottedness. For example the linking number of the flux tubes shown in Fig.~\ref{fig:fig1},inset is 1 whereas the self-linking (Twist+Writhe) of the tube in Fig.~\ref{fig:fig4},inset is 3. 

 $L_{ij}$  can be  computed  by visual inspection, first projecting the tubes onto a plane and subsequently  counting the crossings in an oriented way\cite{Rolfsen:2003p1729}. Given expressions for a pair of  closed lines $\bm{c_1}(\tau)$, $\bm{c_2}(\tau)$,  $L_{12}$ can alternatively be calculated using  the Gauss linking integral:
\cite{Rolfsen:2003p1729,CBaez:1994p1724,Berger:1999p1722}:
\begin{equation}
L(\bm{c_1},\bm{c_2})=\frac{1}{4 \pi} \int  
\frac{d \bm{c_1}}{d \tau_1} \cdot \frac{\bm{c_1}-\bm{c_2}}{\vert \bm{c_1}-\bm{c_2} \vert^3} \times \frac{d \bm{c_2}}{d \tau_2} { \mathrm{d}} \tau_1 {\mathrm{d}} \tau_2,
\label{eq:linking}
\end{equation}

The flux-weighted average linking between tubes added to the self-linking of each tube is then: 
\begin{equation}
K = \sum_{i,j=1}^N L_{ij} \phi_i \phi_j.
\end{equation}
Making the flux tubes finer and finer by letting $N \rightarrow \infty$ and $\phi \rightarrow 0$ it can be shown~\cite{Berger:1999p1722}  that $K$ becomes:
\begin{equation}
K \rightarrow -\frac{1}{4 \pi} \int \int  \mathrm{d}^3 x \  \mathrm{d}^3 y \ \bm{B}(x) \cdot  \frac{\bm{x}-\bm{y}}{\vert \bm{x}-\bm{y} \vert^3 }, \times \bm{B}(y) 
\end{equation}
which using:
\begin{equation}
\bm{A}(x) = -\frac{1}{4 \pi} \int  \mathrm{d}^3 y \    \frac{\bm{x}-\bm{y}}{\vert \bm{x}-\bm{y} \vert^3 } \times \bm{B}(y), 
\end{equation}
becomes the expression for  $h_m$ (Eq.~\ref{eq:eq1}). Note that $\bm{A}(x)$ satisfies $\bm{B}=\bm{\nabla}\times \bm{A}$, and can therefore  be interpreted as the vector potential. 

\begin{figure}[!t]
\begin{center}
\includegraphics[width=0.7 \columnwidth]{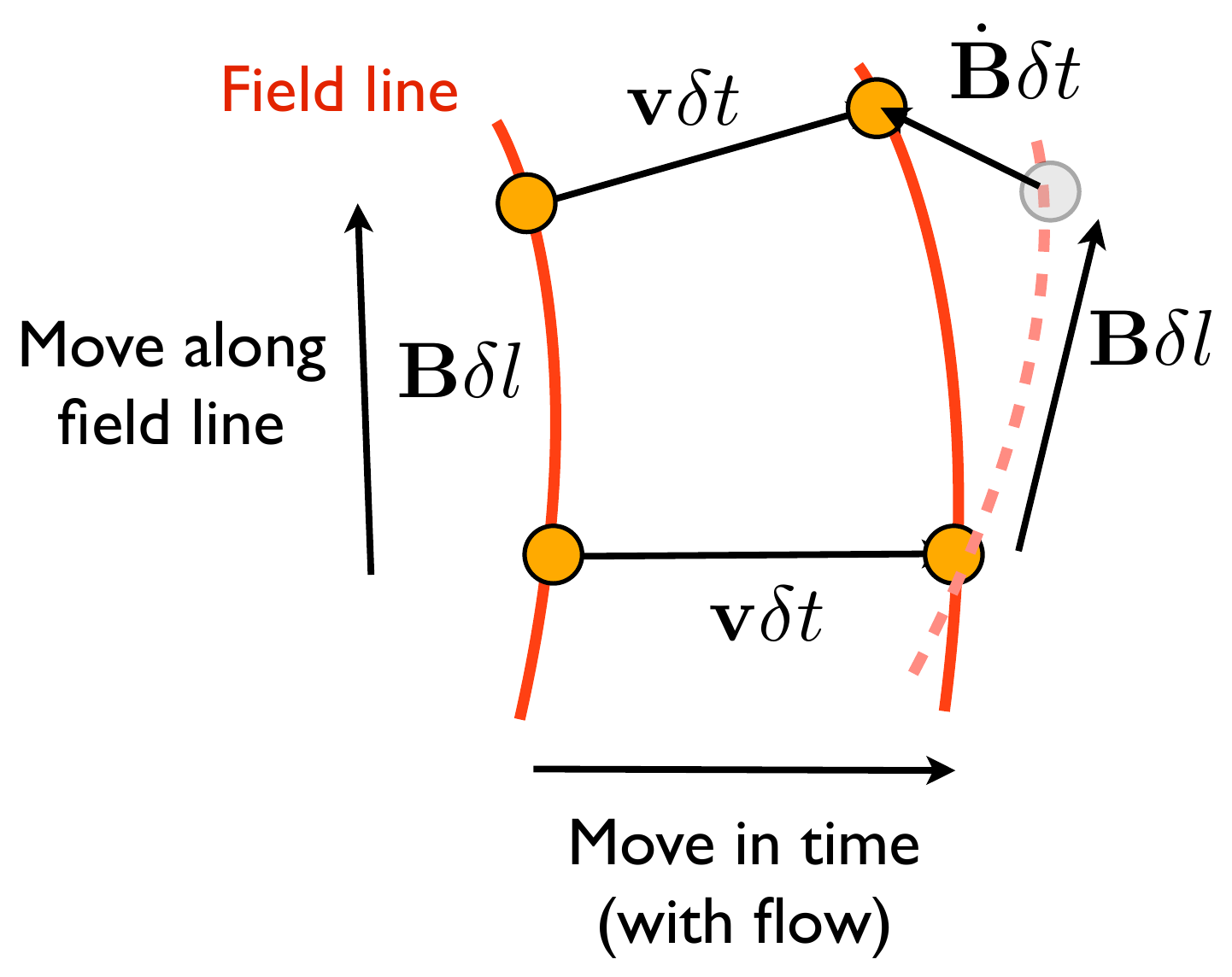}
\caption{For a field line to have a continuous identity in time, it must be possible to to describe its evolution by a velocity field  $\bm{v}$. For this to be true, a field line at time $t_0$, that passes through the point $x_0$, transported by the velocity field $\bm{v}(x)$ must coincide with a field line obtained by adding the field at position $x_0 + \bm{v}(x_0)\delta t$ with the change in the field $\dot{B} \delta t$: $x_0+\delta t \ v_0 +\delta l \ (B_0+\delta t \ (v_0 \cdot \nabla B_0  + \dot{B}_0))=x_0+\delta l \ B_0  + \delta t \ (v_0+\delta l \ B_0\cdot\nabla v_0 )$ or $B_0 \cdot \nabla v_0-v_0 \cdot \nabla B_0=\dot{B}_0$.
This is equivalent to demanding that the commutator of the magnetic field with the velocity field be equal to the time derivative of the magnetic field: the `frozen field' condition. }   
\label{fig:fig2}
\end{center}
\end{figure}
The time derivative of the magnetic helicity is then given by:
\begin{equation}
\partial_t h_m =  \partial_t  \int  \mathrm{d}^3 x \ \bm{A}(x) \cdot \bm{B}(x) \propto \int  \mathrm{d}^3 x \ \bm{E}(x) \cdot \bm{B}(x),
\end{equation}
and magnetic helicity is therefore conserved if $\int  \mathrm{d}^3 x \ \bm{E}(x) \cdot \bm{B}(x) =0$. Similar results hold for the electric helicity.

Since electric/magnetic helicity can be interpreted in this topological manner, it is tempting to associate conservation of electric/magnetic helicity with the preservation   of the topological structure of the electric/magnetic field lines~\cite{Ranada:1989p1769}  however, while this is true {\it on average}, this is  not the case {\it for each field line}.  

\section{The `frozen field' condition}

   \begin{figure}
\includegraphics[width=0.99 \columnwidth]{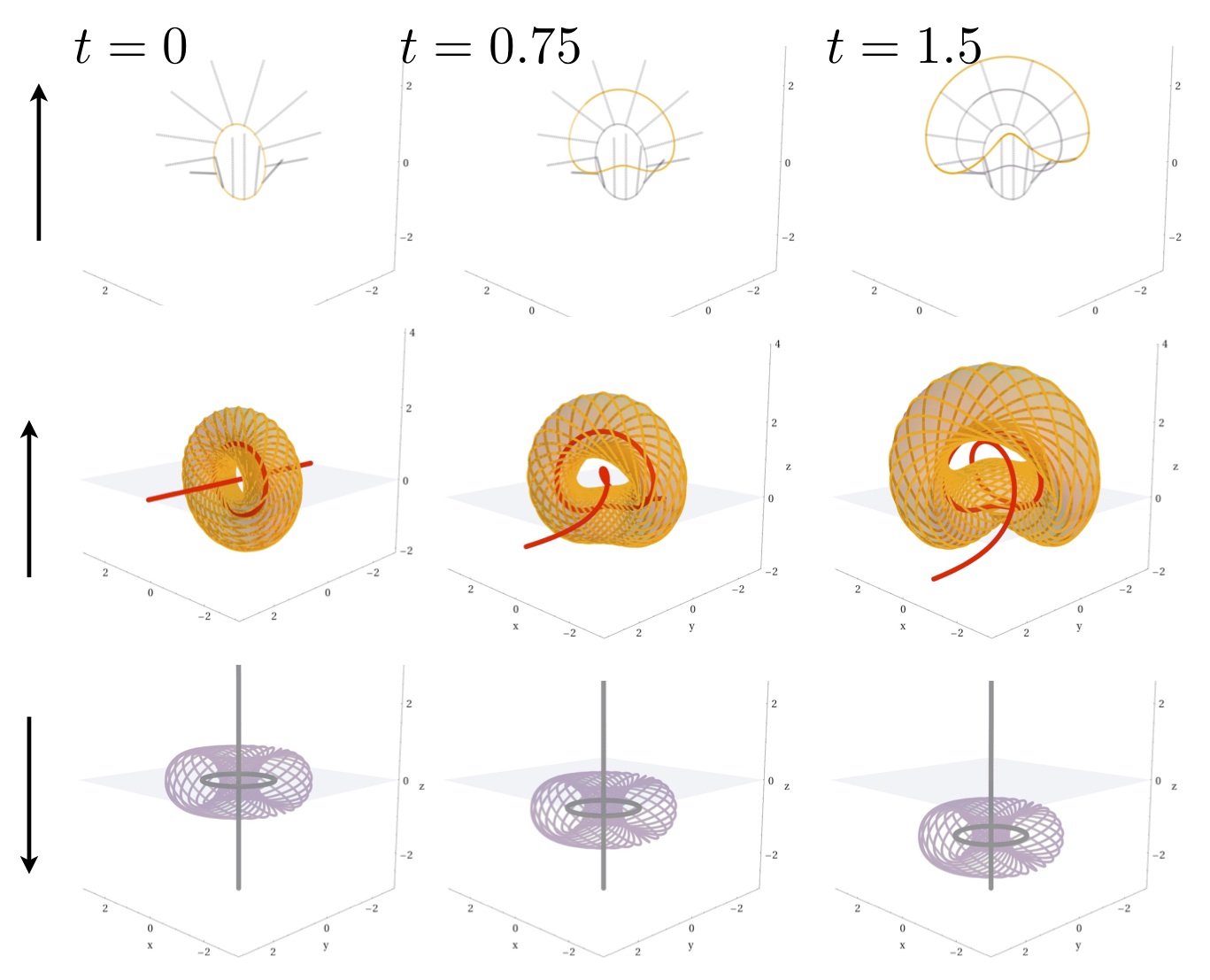}
\caption{Time evolution of the electric and Poynting field lines. The electric field lines (middle) start out as a Hopf fibration oriented along the $x$ axis and evolve by translation in the positive $z$ direction accompanied by some twist and distortion. The evolution of the field lines  is given by  a velocity field whose magnitude is the speed of light and is tangent to the  Poynting vector  field lines (bottom). Unlike the electric field lines, the Poynting field line structure  remains rigid  and  evolves via  a simple translation in a direction  {\it opposite} to the propagation of the field  energy. The trajectory taken by each element of any field line (top) is  a  straight line, tangent to the Poynting field at time $t=0$.   }   
\label{fig:fig3}
\end{figure}

For it to be possible to think of  a field line as an elastic string that cannot be broken and evolves by deformation,   it must be possible to  describe its evolution by a velocity field $\bm{v}$ that transports it. This velocity field must be (a) continuous and (b) such that  the field line transported from one time to another coincides with  a field line at the new time. This latter condition, represented diagrammatically in Fig.~\ref{fig:fig2} amounts to demanding that a field line at time $t_0$, that passes through the point $x_0$, transported by the velocity field $\bm{v}(x)$ must coincide with a field line obtained by adding the field at position $x_0 + \bm{v}(x_0)\delta t$ with the change in the field $\dot{B} \delta t$~\cite{Newcomb:1958p1003}:
\begin{equation}
\left[ \bm{B},\bm{v} \right] - \dot{\bm{B}} = 0,
\label{eq:frozen}
\end{equation}
where $[,]$ denotes the vector field commutator. An equivalent  expression is  
$\bm{\nabla} \times (\bm{v} \times  \bm{B} ) - \dot{\bm{B}} = 0$. This condition is known as the `frozen field' condition.
Taking  $\bm{v}$ to be flux-preserving 
$\bm{v} = (\bm{E} \times \bm{B})/(\bm{B}\cdot \bm{B})$, 
the condition reduces to:
\begin{equation}
\bm{\nabla} \times (\bm{v} \times  \bm{B} ) - \dot{\bm{B}} = \bm{\nabla} \times \left( \hat{\bm{B}} (\bm{E}\cdot \hat{\bm{B}}) \right) = 0,
\end{equation}
which is satisfied identically by a field which satisfies $\bm{E}\cdot \bm{B}=0$. Such a field therefore can be thought of as `frozen' into a fluid that flows with a smooth velocity $\bm{v}$. The entire evolution of the field  is therefore encoded in  a smooth, conformal deformation of space and the field lines maintain their identity throughout.

\section{The evolution of a linked null electromagnetic field and its flow field: a congruence of Robinson}

We now  calculate and study the velocity field of our example null solution. In doing so, we will illustrate not only general aspects of the geometric evolution of a null field, but also develop a vivid picture of the geometry of this particular solution.

The velocity field $\bm{v}=(\bm{E}\times \bm{B})/(\bm{B}\cdot \bm{B})$ for the Hopf electromagnetic knot of Figure~1, is given by: 
\begin{equation}
\bm{v} = \frac{1}{(1 + (t+z)^2 + x^2+y^2  )}
\left(
\begin{array}{c}
 2 ( x (t+z)  -y) \\
 2 (y (t+z)+x) \\
1+ (t+z)^2  -x^2-y^2
\end{array}
\right).   
   \end{equation}
 At time $t=0$, $\bm{v}$ is tangent to a Hopf fibration aligned with the $z$ axis (Fig.~\ref{fig:fig3}). Remarkably,  from its functional dependence on of $z-t$,   $\bm{v}$ can be immediately seen to evolve by simply moving without change of form along $-z$ (opposite to the energy propagation direction) at the speed of light (See Fig.~\ref{fig:fig3}). Such a structure is known in the literature as the Robinson congruence~\cite{robcon}.
 
Even though the instantaneous velocity field has this intricate, structure,  the trajectory of a single element of a field line is remarkably simple: each element travels along a   straight line that was tangent to the flow at $t=0$. This is  illustrated in Fig.~\ref{fig:fig3} in which the straight line trajectories taken by the elements of a field line are shown, with the field line overlaid at different times. It is easy to verify  by explicit computation that the field line propagated  along such straight paths, at the speed of light,  to a new time,  is indeed a field line at this new time.  In oder to determine the field at a later time, it is therefore not  necessary to know the form of the velocity field at all times, but only at $t=0$.

The null condition which ensures the existence of such a propagation by deformation: $\bm{E}\cdot \bm{B}=0$ is also satisfied by plane waves, but  is not satisfied in general by  electromagnetic fields in free space. In particular it is not even a conserved quantity: fields that satisfy  $\bm{E}\cdot \bm{B}=0$ at time $t=0$, do not in general satisfy $\bm{E}\cdot \bm{B}=0$ at later times. 
 For null fields such as  the case under consideration,  the preservation of  $\bm{E}\cdot \bm{B}=0$ is guaranteed if the propagation induced by the velocity field  is free of shear~\cite{ROBINSON:1961p1773,Bampi:1978p1782} (so that the angle between $\bm{E}$ and $\bm{B}$ does not change). In the language of Robinson,  the straight lines along which the field lines propagate correspond to the geodesics of a geodesic shear free congruence.

Before moving on to examine the role of electric/magnetic helicity conservation in the evolution of the field, we note that in the topological model written of Ra\~nada, the orthogonality condition is also guaranteed within the model if admissable Cauchy data is chosen for the initial value problem\cite{Ranada:1989p1769}. 
Finally we note in passing that as pointed out in Ref.~\cite{BialynickiBirula:2004p1740}, the solution  can be obtained from a complex conformal transformation (inversion) on a circularly plane wave, in a way perhaps analogous to the method of generating such a solution proposed in Ref.~\cite{Irvine:2008p1699} by focussing down of a circularly polarized beam.

\section{The evolution of a knotted non-null field with conserved magnetic helicity}
\begin{figure}
\includegraphics[width=0.99 \columnwidth]{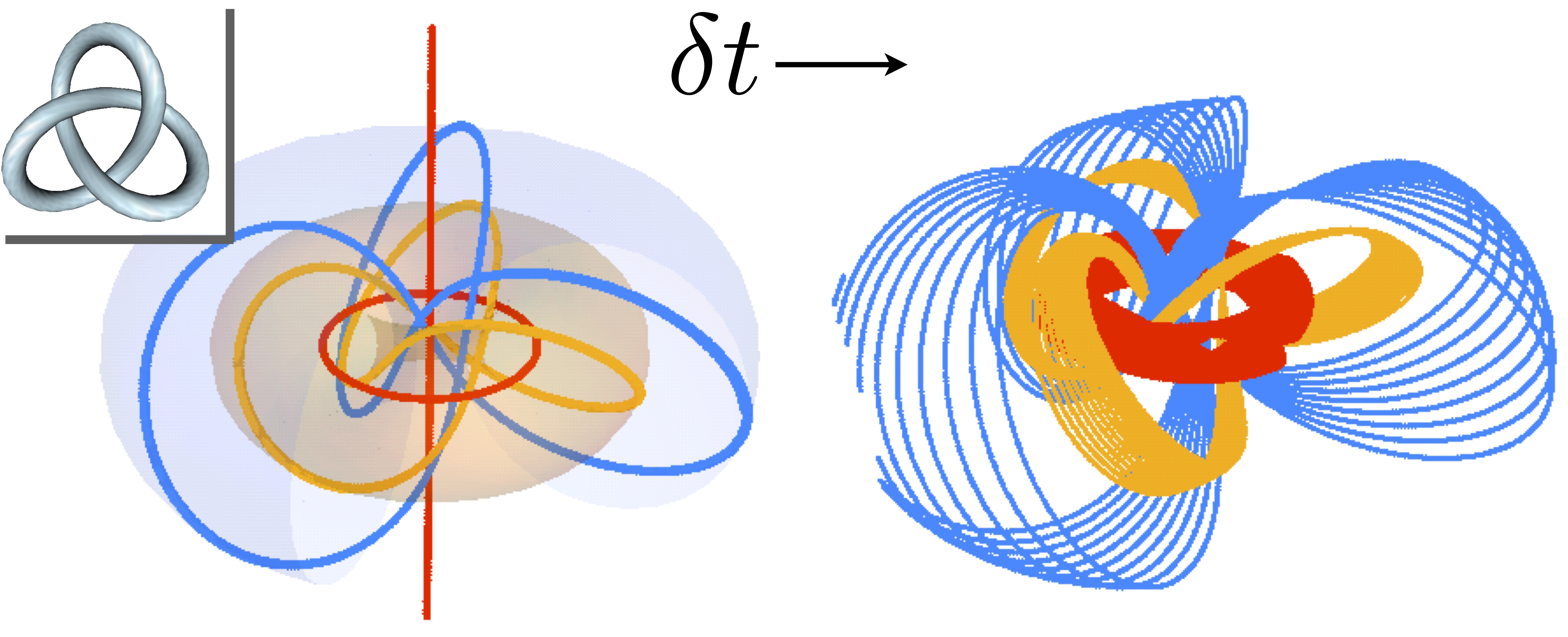}
\caption{Magnetic field lines of a  solution whose magnetic helicity is conserved, but whose electric and magnetic field are not everywhere perpendicular. The initial structure of the field is that of an infinite set of  trefoil knots. Although the average measure of linking and knottedness as measured by the magnetic helicity is conserved, the field lines unravel and do not have a continuous identity in time, providing a striking example of the difference between conservation of magnetic/electric helicity and evolution by smooth deformation. }   
\label{fig:fig4}
\end{figure}
Nullness will guarantee continuity of the field lines. To  clarify the different roles played by the orthogonality condition on $\bm{E}$ and $\bm{B}$, and  magnetic/electric helicity conservation, we now consider an example  in which the field is not null and magnetic helicity is conserved ($\bm{E}\cdot \bm{B}\neq 0, \int  \mathrm{d}^3 x \ \bm{E}(x) \cdot \bm{B}(x) = 0$). 

This example was constructed using a decomposition of the field into vector spherical harmonics found by the author in Ref.~\cite{Irvine:2008p1699}. In this picture the decomposition of the field studied so far is given by:
\begin{equation}
\bm{A}(\bm{r},t) =  
 \int {\rm d}k \ 
 k^3 e^{-k} 
\bm{A}^{-}_{1,1}(k,\bm{r}) e^{-i\omega t} + c.c.,
\label{eq:sphspec2}
\end{equation}
where $\bm{A}^{-}_{1,1}(k,\bm{r}) $  is   Chandrasekhar-Kendall~\cite{Chandrasekhar:1957p1866} curl and angular momentum eigenstate
 that satisfies $\bm{\nabla} \times \bm{A}^{\pm}_{l,m}(k,\bm{r}) =  \pm k \bm{A}^{\pm}_{l,m}(k,\bm{r})$~\footnote{ 
$\bm{A}^{\pm}_{1,1}(k,\bm{r}) = \left[ \bm{A}^{TE}_{1,1}(k,\bm{r}) \pm i  \bm{A}^{TM}_{1,1}(k,\bm{r}) \right] $, $\bm{A}^{TE}_{lm}(k,\bm{r}) = \frac{1}{i\omega} f_l(kr) \bm{L}Y_{lm}(\theta,\phi) $, $\bm{A}^{TM}_{lm}(k,\bm{r}) =  \frac{1}{k^2}\bm{\nabla} \times \Big[f_l(kr) \bm{L} Y_{lm}(\theta,\phi) \Big]$, $\bm{L}=- i \bm{r}\times\bm{\nabla}$ and  $f_l(kr)$ is  a linear combination of the spherical bessel functions $j_l(kr)$, $n_l(kr)$, determined by boundary conditions. In free-space $f_l(kr)=j_l(kr)/\sqrt{l(l+1)}$. }.

An example of  a field which is non-null but whose helicity is  conserved in time  is shown in Fig.~\ref{fig:fig4}. The expression for the field  is:
\begin{equation}
\bm{A}(\bm{r},t) =  \int {\rm d}k \ 
 k^3 e^{-k} e^{-i\omega t} 
 \left[
\bm{A}^{TE}_{1,1}(k,\bm{r}) - i \frac{p}{q}  \bm{A}^{TM}_{1,1}(k,\bm{r})
\right]   + c.c.,
\end{equation}
where $\bm{A}^{TE} = \bm{A}^{+} + \bm{A}^{-}$ and $\bm{A}^{TM} = - i \left(\bm{A}^{+} - \bm{A}^{-}\right)$. 
At time $t=0$, the field lines lie on tori that are identical to those of the Hopf fibration, but on each torus the field lines wrap $q$ times around the azimuthal direction and $p$ times around the poloidal direction. For $p = 2$ and $q = 3$ they correspond to trefoil knots (Figure~\ref{fig:fig4});  for general co-prime $p,q$, they correspond to all possible torus knots~\cite{Rolfsen:2003p1729}. 

For this solution $\bm{E}\cdot \bm{B}$ is non-vanishing, but its spatial average is:
\begin{equation}
\bm{E}\cdot \bm{B} \propto \frac{ r \left(r^2-1-t^2 \right) \cos
   (\theta )}{\pi 
   \left(r^4-2 r^2(t^2-1)+(1+t^2)^2\right)^3},
     \end{equation}
  \begin{equation}
 \int \mathrm{d}^3 x 
\   \bm{E}\cdot \bm{B}  =0.
\end{equation}
The magnetic helicity is therefore conserved. Although magnetic helicity is conserved, 
as the solution evolves,  the magnetic field lines visibly change topology by unraveling as can clearly be  seen in Fig.~\ref{fig:fig3}.

\section{Conclusions}
To investigate the rules that govern the evolution of field lines in knotted beams of light, we have studied the evolution of the field lines of a free-space solution to Maxwell's equations, representing a pulsed beam of light whose electric, magnetic, and Poynting field lines start out as mutually orthogonal Hopf fibrations~\cite{Ranada:1989p1769,Irvine:2008p1699,ROBINSON:1961p1773,Trautman:1977p1705,Besieris:2009p1739,BialynickiBirula:2004p1740}. 
We have shown that the electric and magnetic field lines travel and deform upon propagation as if they were filaments embedded in a fluid flowing along the lines of the Poynting vector at the speed of light. We show that this can be done because the field is null and satisfies the `frozen field' condition.

While the electric and magnetic field lines deform upon evolution, those of the flow field, aligned with the Poynting vector, remain rigidly arranged in the form of a Hopf fibration moving  along the propagation ($z$) axis at the speed of light in the direction {\it opposite} to the energy flow,  a well known structure (the Robinson congruence) in this type of field~\cite{ROBINSON:1961p1773,Trautman:1977p1705,robcon}.  Though the instantaneous flow field has this intricate structure,  each segment of the electric and magnetic field lines  evolves along a straight line tangent to the flow/poynting field at time $t=0$.
 
We have further  examined  a solution in which the magnetic/electric helicity, or average knottedness of the field lines,  is conserved and whose field lines correspond to an infinite set of trefoil knots at time $t=0$.  The electric and magnetic fields in this case however, are not everywhere orthogonal and the field lines break up on evolution.
While the null, shear-free condition on a field does not directly relate to knottedness, it appears that it has a key role to play in the evolution of electromagnetic fields with linked and knotted field lines.

\section{Acknowledgements}

The author thanks Stephen Childress, Mark Dennis, Paul Chaikin, Jan-Willem Dalhuisen and Feraz Azhar for useful conversations. The author acknowledges support from the English-Speaking Union through a Lindemann Fellowship.

\end{document}